\documentclass[12pt]{article}
\begin{document}
\LaTeX
\bigskip
\begin{center}
Theoretical studies of Q1D organic conductors:a personal review 
\end{center}

\medskip
\begin{center}
Vladan \v Celebonovi\'c
\medskip

Institute of Physics,Pregrevica 118,11080 Zemun-Beograd

Serbia and Montenegro
\end{center}
\medskip 
\begin{center}
e-mail:vladan@phy.bg.ac.yu
\end{center} 
\begin{abstract}
The aim of this contribution is to review some aspects of theoretical studies of a family of quasi one-dimensional (Q1D) organic conductors known as the {\it Bechgaard salts}.In order to make it personal,it will retrace the evolution of the
author's interest in this field.

The generalized aim was the calculation of the electrical conductivity of the Bechgaard salts and gaining some knowledge on the equation of state and thermal properties of these materials. The review ends with some ideas about the possible future developement of the field.
\end{abstract}


function method; Hubbard model \\ 

{\it Accepted for publication by NOVA Science Publishers} 
\newpage
\section{Introduction}
I have entered the Institute of Physics (IoP) in Beograd in the spring of 1985. Several years before that I have graduated in physics at the University of Beograd,and after some time spent in looking (in vain) for a job in various research establishements was delighted by the fact that I was finally accepted in IoP. I have entered the Laboratory for interdisciplinary studies,and my research field was broadly defined as " the study of the behaviour of materials under high pressure" . My immediate task was to calculate the electrical conductivity under high static pressure of several {\it normal} metals.If I suceed in the calculation,my job in IoP would became stable.If not.... My first contact with the organic conductors dates from that hectic period,when I discovered the existence of these curious materials in the literature.

My real encounter with the organic conductors occured in Paris in France a few years after that.In the spring of 1987.,I was awarded a French governement schollarship for a visit of 8 months to  Laboratoire des interactions moleculaires et hautes pressions (LIMHP) at Universit\'e Paris-Nord.The aim of my visit was to learn experimental techniques of high pressure research. 

While working in LIMHP,I managed (through a friend of a friend from Belgrade) to arrange a visit to Laboratoire de Physique des Solides du CNRS on the campus of Universit\'e Paris $XI$ in Orsay.I was to be received by "someone" named Denis J\'erome. Even while I am writing this,after nearly 17 years,I am smiling when I think that on the day when my visit was being fixed,this name did not mean anything special to me! However,the afternoon before I was due to go to Orsay,I received a reprint of a then recent paper from "Contemporary Physics" from which I learnt that the man whom I was going to meet actually discovered most of the facts related to quasi one-dimensional (Q1D) organic conductors.

The rest of the story is "condensed history". The following day I went to Orsay and met D.J\'erome.He received me extremely kindly,we had a friendly conversation,focused on high pressure and organic conductors ,and  somewhere around noon we went to the canteen. While there,I started thinking that it would be "great" if I could somehow get the possibility to work for my Ph.D.in J\'erome's group.However,I was in a dilemma how to ask him such a thing after meeting him for the first time in my life.Whether or not he felt my thoughts I do not know.The important thing happened when in the course of the lunch we came to delicious camembert cheese.
\newpage
J\'erome simply asked me whether I would be interested to come and work some time in his laboratory.I started laughing,explained him my toughts,and..
that was it.We agreed that I would come to Orsay and work for my Ph.D.there but that I would defend it in Belgrade. When my schollarship ended,I went back to Belgrade,completed and defended my M.Sc.,prepared all the necessary administrative formalities,and in autumn of 1989.,I was in Paris again,with a schollarship for 12 months. 

J\'erome's group was (and still is) one of the leading centers of studies of the organic metals,in particular of the Bechgaard salts.The name of these materials is derived from the fact that they were synthetized by K.Bechgaard and his group in Denmark \cite{BECH:80},and their generic chemical formula is $(TMTSF)_{2}X$.The symbol $(TMTSF)_{2}$ denotes a complicated molecule of the name of di-tetra-meta-tetra-selena-fulvalene,
which is the basic ingredient of all the Bechgaard salts,and $X$ denotes an anion such as $PF_{6},FSO_{3},NO_{3}...$.

When the Bechgaard salts were discovered,it was thought that they mar
ked "the road" to be taken in the quest for high temperature superconductivity. This turned out to be wrong, so the motivation for their study (both experimental and theoretical) somewhat changed.The main motivation for their study nowadays is the idea that they are the simplest possible cases (and therefore easier to understand) of correlated electron systems,
which (it is thought worldwide) the high $T_{c}$ materials are. 

Time went quickly and I worked on the experiment aimed at measuring the electrical conductivity of the Bechgaard salt $(TMTSF)_{2}FSO_{3}$ \cite{PAVL:91}. When the experiment was almost complete,and we were starting to plan the continuation of it with the introduction of impurities,a critical part of the experimental setup exploded.No one was harmed,but waiting for the replacement I started trying to do some theoretical work on the conductivity of the Bechgaard salts.J\'erome noted it and put me into contact with Heinz J.Schulz of the theory group.A couple of months before that,on advice of J\'erome I have asked for the prolongation of my schollarship for another year. This was at first rejected,but the Laboratory (and J\'erome personally) used their "contacts" in the administration.After the summer holidays,when I got back to Paris,I found a hand written note from J\'erome saying "Dear Vladan,
wellcome for another year in Paris".I was proud and happy,passed to the theory group to work with Schulz,and that is where my theoretical work  on the organic conductors begun.
\section{The calculation of the conductivity}
\subsection{The first attempt} 
Heinz Schulz was a tall, quiet, friendly man spending all the day in the laboratory,except when he had  courses at the Ecole Politechnique.He was from Hamburg,got his Ph.D.there and came to Orsay as a post-doc.He prepared another Ph.D. there,and was immediately offered a job.Sadly,he died of cancer in 1996.,and only then I discovered that we were of nearly the same age.We agreed that after my experimental work in J\'erome's group it was logical that I continue by trying to calculate the electrical conductivity of the Bechgaard salts. 

Two immediate questions to which we had to find answers were the choice of the method of calculation,and the choice of the Hamiltonian (i.e.,a theoretical model of the material)  which could be an appropriate description of the Bechgaard salts.

The logical choice of the theoretical model of these materials was the famous Hubbard model.It was proposed several decades before the start of my work in Orsay,and it contains the basic "ingredients" for the correct theoretical description of a solid:atoms,localized in the nodes of some form of a crystal lattice and electrons localized on them \cite{HUBB:63}.Although it is intuitively simple and acceptable,to this day the Hubbard  model was solved only in the simplest(one-dimensional) case \cite{LIWU:68}.In one spatial dimension the Hamiltonian of the Hubbard model has the following form:

\begin{equation}
	H=-t\sum_{i,\sigma=0}^{N}(c_{i+1,\sigma}^{+}c_{i,\sigma}+c_{i,\sigma}^{+}c_{i+1,\sigma})+U\sum_{l=0}n_{l,\uparrow}n_{l,\downarrow}=H_{0}+H_{1}
\end{equation}

The first term is the kinetic energy term $(H_{0})$,and the
second one is the interaction term $(H_{1})$.
All the symbols in this equation have their standard meanings: $t$ is the electron hopping energy,the terms in parenthesis in $H_{0}$ are the creation and anihillation operators for electrons on a lattice site $i$ with spin $\sigma$.In the interaction term,$U$ is the interaction energy of a pair of electrons having oposite spins within an atom on lattice site $l$.

A starting idea for the calculation of the conductivity was to try and apply the historic Kubo formula \cite{KUB:57}.In modern transcription,such as \cite{WABA:86}, Kubo's formalism gives the following expression for the real part of the conductivity tensor:
\begin{equation}
\sigma^{'}_{\mu\nu}(\omega)= \frac{\omega}{2 V \hbar} (1-exp^{-\beta\hbar\omega}) I 	
\end{equation}

and $I$ denotes the following 

\begin{equation}
I=\int_{-\infty}^{\infty} exp^{it \omega} \left\langle M_{\mu}(t)M_{\nu}(0)\right\rangle_{0}
\end{equation} 
where $M_{\mu}$ is the time dependent dipole moment operator.In the Heisenberg picture,this time dependence is expressed as 

\begin{equation}
M_{\mu}(t)=exp^{itH_{0}/\hbar}M_{\mu}exp^{-itH_{0}/\hbar}
\end{equation}
\newpage
To complicate things further,the calculation of the correlation function in the expression for $I$ demands the knowledge of the partition function and the density matrix of the system. Obviously,the Kubo formula although being exact is too abstract for performing a realistic calculation. 

\subsection{The memory function approach}

After we jointly concluded that trying to apply Kubo's formula  was actually leding to an {\it impasse} Schulz suggested that I try applying the so called "memory function" method.Of course I accepted,and had an immediate feeling that I was entering a sort of scientific "terra incognita". 

The memory function method has its roots in work in pure rigorous statistical mechanics around the middle of the last century (\cite{MO:33},\cite{MO:34} and earlier work cited there). A theory of many-particle systems was developed in these papers with the aim of describing transport properties,collective motion and Brownian motion from a statistical-mechanical point of view.This formulation,taken from the abstract of \cite{MO:33}, does not sound complicated.However,retracing in detail the calculation performed by Mori is a task of considerable mathematical complexity. The main point of using the memory function in any calculation of transport properties is that it gives the possibility of evading the problem of formulating and solving the transport equations.Instead of dealing with the transport equations,this method gives the possibility of expressing the conductivity in terms of a regular "memory function".

Papers \cite{MO:33} and \cite{MO:34} are,in some sense,comparable to \cite{KUB:57}:they are exact in principle,but applicable
to real calculations with extreme difficulty.However,
some time after the publications of Mori,a technique 
was worked out which gave the possibility of applying 
the memory function method to real calculations 
\cite{GW:71},\cite{GW:72}.

The equations needed for the calculation of the conductivity  within the memory function method are \cite{GW:72}
\begin{equation}
	\chi_{AB}(z)=<<A;B>>=-i\int_{0}^{\infty}exp^{(izt)}<[A(t),B(0)]>dt
\end{equation}
and
\begin{equation}
	\sigma(\omega)=i(\frac{\omega_{P}^{2}}{4z\pi})[1-\frac{\chi_(z)}{\chi_{0}}]
\end{equation}

\newpage
In these equations $\omega_{P}^{2}=4\pi n_{e}e^{2}/m_{e}$ 
is the square of the plasma frequency,
$n_{e}$,
$e$ and $m_{e}$ 
are the electron number density,charge and mass,while 
$\chi_{0}$ denotes the ratio $n_{e}/m_{e}$,which  
is the zero frequency limit 
of the dynamical susceptibility $\chi(\omega)$.  

The integral in eq.(5) is a general definition of the linear response of an operator $A$ to a perturbing operator $B$. It is an analytic function for all non-real frequencies $z$ \cite{GW:72}.The symbol $A(t)$ denotes the Heisenberg representation of the operator A.Eq.(5) can "evolve" from a general definition of a response function into the definition of a current-current correlation function by introducing $A=B=[j,H]$.The current operator is denoted by $j$ and $H$ is the Hamiltonian of the system under consideration.

It was decided,in line with \cite{AND:87}, to use the Hamiltonian of the Hubbard model in the calculation.The Hamiltonian of this model \cite{HUBB:63} in one dimension is given in eq.(1). The current operator has the form

\begin{equation}
	j=-it\sum_{l,\sigma}(c_{l,\sigma}^{+}c_{l+1,\sigma}-c_{l+1,\sigma}^{+}c_{l,\sigma})
\end{equation}

The final step which needs to be made before the actual calculation of the conductivity of the Bechgaard salts is the determination of the chemical potential of the electron gas on a 1D lattice.It may not be obvious at first why is the chemical potential so important for this work.The usual research practice is that pure Bechgaard salts are described as systems with $\mu=0$.Any possible doping (and experiments with doped specimen are very interesting) is theoretically described by deviating $\mu$ from 0.The chemical potential is a function of the band-filling $n$,the inverse temperature $\beta$ and the electron hopping energy $t$.A determination of an analytical expression for this function demands a choice of a theoretical model for the description of the electrons in the Bechgaard salts: Fermi or Luttinger liquid? 
\subsubsection{Fermi or Luttinger liquid?}

The basic ideas of what is now (for several decades already) called the Fermi-liquid theory are present in physics for a long time - it can be safely stated since the time of Sommerfeld. He showed that various experimental data on metals (for example the low temperature behaviour of their specific heat or the electrical conductivity) can be understood by assuming that the electrons in a metal behave like a gas of non-interacting Fermions.Shortly after Sommerfeld,came the results of Pauli,Bloch and Wigner.
\newpage
They showed that the paramagnetic susceptibility of the non-interacting electrons is temperature independent,and that the interaction energies of the electrons at metallic densities are comparable to their kinetic energy.The 
"final touch" was given by Landau \cite{LALA:56},\cite{LALA:57}.In these works Landau introduced a new way of thinking about interacting systems,which became crucial for the developement of condensed matter physics.For example,the notions of quasiparticles and elementary excitations,were introduced in these papers.Methodologically,they were extremely important,because in them Landau introduced the idea of "asking useful questions about the low energy excitations of the system based on concepts of symmetry,without worrying about the myriad of unnecessary details" \cite{VAR:02}.

When does the Fermi liquid theory break down? Physically,the answer is logical: it happens whenever one of the measurable quantities which it aims to calculate diverges \cite{VAR:02}.It has been shown in field theory that the Fermi liquid concept breaks down in one dimension. Technically speaking,some vertices which are in the Fermi liquid theory assumed finite,diverge in one dimension because of the Peierls effect (for example \cite{VO:00}).It is also known that the excitations in 1D are not quasi particles,but collective charge and spin degrees of freedom,each of which propagates with different velocity \cite{VO:93}.

Optical and photoemission experiments on the Bechgaard salts show deviations from the Fermi liquid behaviour \cite{VE:00}. However,the energy scales on which the deviations occur in these two groups of experimental data are different.Interesting conclusions have been reached from c-axis conductivity measurements of the Bechgaard salts \cite{MO:98}.It was shown there that the 1D Luttinger liquid description of the Bechgaard salts breaks down below a pressure dependent value of the temperature,and that the Fermi liquid description is restored at low temperatures of the order of $10$ K.

This subsection is certainly not a complete review of the problem "Fermi vs.Luttinger".It's aim is to try to justify to the reader the choice which was made in this work: to use the Fermi liquid theory.The logical question is "Why?". 

There are several motives for this decision.First and simplest - chronologically,this calculation of the conductivity of the Bechgaard salts was completed in 1997.and started while the present author still was in Orsay.Second,at the time when this calculation was being performed many experimental data which are now avaliable simply did not exist.
\newpage
This implied that it is wiser to perform the calculation within the Fermi liquid theory as a very well known theoretical framework.Experimentally,it was known that the Bechgaard salts are not strictly 1D but quasi 1D - the conductivity along the so called c-axis is not 
exactly equal to zero.There is also the 
{\it aposteriori} justification: results for the temperature
dependence of the resistivity are qualitatively similar to experimental data.

\section{The calculation-the practical part }
\subsection{The chemical potential} 
It is known since \cite{LIWU:68} that the chemical potential of the electron gas is zero for the single-band half-filled Hubbard model.The band filling is defined as the rato of the number of electrons present per lattice site $n_{e}$ to the maximal possible number of electrons per site $N$. This value is obviously equal to 2 since two electrons with different spins can be present on the same lattice site.
\begin{equation}
	n=\frac{n_{e}}{N} 
\end{equation} 

Theoretically speaking,the value of $n_{e}$ is a parameter which can be varied at will.From the experimental viewpoint,this value can be changed by doping the specimen with electron donors or acceptors.As the chemical potential enters the expression for the Fermi fermi function,the first practical step in calculating th electrical conductivity had to be the determination of the chemical potential of the electron gas on a 1D lattice \cite{CHE:96}.

The starting point of this calculation is the following equation:

\begin{equation}
	n=\frac{1}{\pi\hbar}\int\frac{1}{1+exp\beta(\epsilon_{k}-\mu)}dp
\end{equation}
where all the symbols have their usual meaning,$p=\hbar k$ and $n$ is the number of electrons per site.The integration is performed within the limits $\pm\pi\hbar/s$ and $s$ is the lattice constant.In the 1D Hubbard model $\epsilon_{k}=-2t\cos(ks)$,which leads to 

\begin{equation}
dp=\hbar dk=\frac{\hbar}{s}(4t^{2}-\epsilon^{2})^{-1/2}d\epsilon
\end{equation}
\newpage
Inserting eq.(10) into eq.(9) gives  

\begin{equation}
n=\frac{1}{\pi s}\int_{-2 t}^{2 t} \frac{1}{1+exp\beta(\epsilon_{k}-\mu)}(4t^{2}-\epsilon^{2})^{-1/2}d\epsilon 
\end{equation}

The initial problem of determining the function $\mu=\mu(\beta,t,n)$ has thus been reduced to the problem of solving the integral equation eq.(11).Instead of proceeding with the full rigour of the theory of integral equations,one can tackle the problem by using a suitable power series representation of the Fermi function. 

The Fermi function can be represented as 

\begin{equation}
\frac{1}{1+exp\beta(\epsilon_{k}-\mu)}=\Theta(\mu-\epsilon)-\sum_{k=0}^{\infty}A_{2k+1}\beta^{-2(k+1)}\delta^{2k+1}(\epsilon-\mu)	
\end{equation}

where the symbol $A_{2k+1}$ denotes

\begin{equation}
A_{2k+1}=\frac{2(-1)^{k+2}\pi^{2k+2}(2^{2k+1}-1)}{(2k+2)!} B_{2k+2}	
\end{equation}
$\Theta$ is the step function,$\delta^{2k+1}(\epsilon-\mu)$ are the derivatives of the $\delta$ function and $B_{2k+2}$ are the Bernoulli numbers.

Inserting eq.(12) into eq.(11) gives

\begin{equation}
n=\frac{1}{\pi s}\int_{-2 t}^{2 t}\left[\Theta(\mu-\epsilon)-\sum_{k=0}^{\infty}A_{2k+1}\beta^{-2(k+1}\delta^{2k+1}(\epsilon-\mu)\right](4t^2-\epsilon^2)^{-1/2} d\epsilon
\end{equation}
Eq.(14) is integrable by using the following relation,valid for any function $f(x)$ and its $n$-th order derivative:
\begin{equation}
\int\delta^{n}(x-x_{0})f(x)dx=(-1)^{n}f^{(n)}(x_{0})	
\end{equation}
 
Inserting eq.(15) in eq.(14) and limiting the summation to terms with $k\leq2$ leads to an equation of second degree in $\mu$ which can be solved to give the following result for the chemical potential of the electron gas on a 1D lattice:

\begin{equation}
\mu=\frac{(\beta t)^{6}(ns-1)\left|t\right|}{1.1029+.1694(\beta t)^{2}+.0654(\beta t)^{4}}	
\end{equation}
\newpage
Obviously,for $lim_{n,s\rightarrow1}\mu$=0,which means that this result has as its limiting value the well known result of Lieb and Wu \cite{LIWU:68}.  

The reader may at this point be tempted to ask for an explanation of the limitation to terms having $k\leq2$ in the calculation leading to eq.(16).The explanation is that the idea being to obtain an analytical expression for the function $\mu(\beta,t,n)$ going to terms with $k\geq2$ leads to expression which are too complicated for any applications in further work. 
 
\subsection{The electrical conductivity}
 
With the chemical potential being known,it finally became possible to tackle the problem of the electrical conductivity of the Bechgaard salts \cite{CHE:97}. The Hubbard hamiltonian and the current operator are given by eqs.(1) and (7).The functions $\chi(z)$ and $\sigma(z)$ are both expressible in complex form as $\sigma(z)=\sigma_{R}(z)+i\sigma_{I}(z)$ and $\chi(z)=\chi_{R}(z)+i\chi_{I}(z)$.If one further assumes that the frequency is a complex function,with $z=z_{1}+i z_{2}$ and that $z_{2}=\alpha z_{1}$ with $\alpha \succ 0$,it follows from eq.(6) that 

\begin{equation}
\sigma_{R}+i\sigma_{I}=i\frac{\omega_{P}^{2}}{4z_{1}\pi(1+i\alpha)}\left[1-\frac{\chi_{R}(z)+i\chi_{I}(z)}{\chi_{0}}\right]
\end{equation}
This expression can be separated into the real and immaginary part.Taking the special case $\alpha=0$ (justificed by the fact that the frequency measured in experiments is real) gives:

\begin{equation}
	\sigma_{R}=\frac{\omega_{P}^{2}\chi_{I}}{4z_{1}\pi\chi{0}} 
\end{equation}
and
\begin{equation}
	\sigma_{I}=\frac{\omega_{P}^{2}}{4\pi z_{1}}\left(1-\frac{\chi_{R}}{\chi_{0}}\right)
\end{equation}

The ensuing steps in the calculation of the electrical conductivity are in principle straightforward.The expression {\it "in principle"} is here amply justified,because inserting all the "`sub-results"' into the expression for $\chi_{AB}$ leads to an almost intractable expression.

The "practical" calculation (practical is here employed in the sense leading to the function $\sigma_{R}$) starts by determining the current-current correlation function.This can be obtained by inserting $A=B=\left[j,H\right]$ into eq.(5).
\newpage
Evaluating this commutator is facilitated to some extent by the decomposition of the the Hamiltonian of the Hubbard model indicated in eq.(1).
Relatively easily it can be shown that $\left[j,H_{0}\right]=0$ and that

\begin{equation}
A=\left[j,H\right]=\left[j,H_{1}\right]=-itU\sum_{l,\sigma}\left((c_{l,\sigma}^{+}c_{l+1,\sigma}-c_{l+1,\sigma}^{+}c_{l,\sigma}\right)\left(\delta_{l+1,j}-\delta_{l,j}\right)n_{j,-\sigma}
\end{equation} 
All the symbols have their standard meanings.Transition to $k$ space can be performed by relations of the following form \cite{EM:79}
\begin{equation}
c_{l,\sigma}^{+}=N^{-1/2}\sum_{k1}\exp(i k_{1}l s)c_{k_{1},\sigma}^{+}	
\end{equation}
In this expression $N$ is the number of lattice sites,$s$ is th elattice constant and $L=N s$ is the length of the specimen.The temporal evolution of various operators can be introduced in the calculations by relation of the form $c_{k}(t)=\exp{-i\epsilon(k)t} c_{k}$ where (for the one dimensional Hubbard model) $\epsilon(k)=-2 t cos(ks)$.Inserting all this into the expression for $\chi_{AB}$ gives the following expression for the susceptibility $\chi(z)$
\begin{eqnarray}
\chi(z)=\sum_{p,g,k,q}(32i(1/[(1+\exp(\beta(-\mu-2tcos(g))))\nonumber\\
(1+\exp(\beta(-\mu-2tcos(k))))]-1/(1+\exp(\beta(-\mu-2tcos(p))))\nonumber\\
1/(1+\exp(\beta(-\mu-2tcos(q))))](Ut)^{2}(\alpha z_{1}+i(z_{1}+2t(cos(q)+cos(p)\nonumber\\
-cos(g)-cos(k))))(cos(p+g)/2)(cos((q+k)/2)[cosh(g-p)-1]/\nonumber\\
(N^{4}((\alpha z_{1})^{2}+(z_{1}+2t(cos(q)+cos(p)-cos(g)-cos(k))))^{2}))
\end{eqnarray}
In order to simplify the calculation at least to some extent,the summation was limited to the first Brillouin zone,and the lattice constant $s$ was set as $s=1$.This obviously implies that $L=N$.The sum in the equation above were calculated under the condition $\alpha\neq0$ because this condition is built in in the definition of $\chi(z)$.Once the sums were calculated,taking into account that in reality the frequency is a real quantity,the limit  $\alpha\rightarrow0$ was imposed. 
After performing all the summations and taking $\alpha\rightarrow0$ the following approximation for the dynamical susceptibility was obtained.
\newpage
\begin{eqnarray}
\chi\cong(32i(-1+cosh(1))[1+exp(\beta(-\mu+2t\cos(1-\pi))))^{-2}\nonumber\\
(1+\exp(\beta(-\mu-2t)))^{-2}](Ut/N^{2})^{2}\cos^{2}((1-2\pi)/2)\nonumber\\
(z_{2}+i(z_{1}+2t(-2-2cos(1-\pi))))
/(z_{2}^{2}+(z_{1}+2t(-2-2cos(1-\pi)))^{2}
\nonumber\\
+<<2267>> 
\end{eqnarray}
	
The number $<<2267>>$ in the preceeding equation denotes the number of omitted terms.Obviously,such a long equation is inapplicable and has to be truncated after a certain number of terms.Taking the first 32 terms of eq.(23),multiplying out the products and powers,expressing the result as a sum,gives the real part of the dynamical susceptibility as

\begin{eqnarray}
\chi_{R}(z)\cong[128U^{2}t^{3}cos^{2}((1-2\pi)/2)]/[1+\exp(\beta(-\mu-2t)))^{2}N^{4}(z_{2}^{2}+
\nonumber\\
+(z_{1}+2bt)^{2})]
+<<527>>	
\end{eqnarray}

and $b=-4 (1+cos(1-\pi))$.
 
Taking the first 20 terms of eq.(24) and imposing the condition $\lim z_{2}\rightarrow0$ it follows that the real component of the function $\chi_{R}$ has the form

\begin{equation}
\chi_{R}(\omega)=\sum_{i}\frac{K_{i}}{(\omega+b t)^{2}}+\sum_{j}\frac{L_{j}\omega}{(\omega+b t)^{2}}	
\end{equation}

The functions $K_{i}$ and $L_{j}$ can be read-off from eq.(24) developed to a given number of terms.Using this result,the immaginary part of the dynamical susceptibility $(\chi_{I})$ is given by \cite{WABA:86}

\begin{equation}
\chi_{I}(\omega)=-2\frac{\omega_{0}}{\pi}P\int\frac{\chi_{R}(\omega)}{(\omega^{2}-\omega_{0})^{2}}	
\end{equation}
Inserting eq.(25) into eq.(26) and imposing the
constraint  $i,j\leq4$ in eq.(25)
leads to the following expression for $\chi_{I}$
\begin{eqnarray}
\chi_{I}=(2bt/\pi)(Ut/N^{2})^{2}[\omega_{0}/(\omega_{0}+2 b t)(\omega_{0}^{2}-(2 b t)^{2})][4.53316(1+
\nonumber\\
\exp(\beta(-\mu-2t)))^{-2}+24.6448(1+\exp(\beta(-\mu+2tcos(1-\pi)))))^{-2}]+
\nonumber\\
(2/\pi)[\omega_{0}/(\omega_{0}^{2}-(2 b t)^{2})](Ut/N^{2})^{2}\times[42.49916(1+\exp(\beta(-\mu-2t)))^{-2}+
\nonumber\\
78.2557(1+\exp(\beta(-\mu+2t\cos(1-\pi))))^{-2}]
\nonumber\\	
\end{eqnarray}
\newpage
The final expression for the electrical conductivity follows by inserting this result into eq.(18).After some algebra,one gets the 
final result for the electrical conductivity of Q1D organic metals
\begin{equation}
\sigma_{R}(\omega_{0})=(1/2\chi_{0})(\omega_{P}^{2}/\pi)[\omega_{0}^{2}-(bt)^{2}]^{-1}(Ut/N^{2})^{2}\times S
\end{equation}

and the symbol $S$ denotes the following function

\begin{eqnarray}
S=42.49916\times(1+\exp(\beta(-\mu-2t)))^{-2}+78.2557\times
\nonumber\\
(1+\exp(\beta(-\mu+2t\cos(1+\pi))))^{-2}+(bt/(\omega_{0}+bt))\times
\nonumber\\	
(4.53316\times(1+\exp(\beta(-\mu-2t)))^{-2}+
\nonumber\\
24.6448(1+\exp(\beta(-\mu+2t\cos(1+\pi)))))^{-2})
\nonumber\\
\end{eqnarray}

\subsection{Discussion}

Equations (28) and (29) are the final result for th electrical conductivity of the Q1D organic metals obtained within the memory function theory.From the purely mathematical point of view they are "moderately complicated",the main "`complicating factor" being the length of the two expressions.However,for applications in physics,the relevant question is whether or not the results of these two expressions agree with experiments.

The main externally controllable parameters in experiments on organic conductors are the temperature,the doping and the frequency,and in the remainder of this chapter,the comparisons of the results of eqs.(28) and (29) will be performed with respect to these parameters. Model parameters which are constant in these two expressions were chosen as follows:$N=150$,$U=4t$,$\omega_{P}=3U$,$\chi_{0}=1/3$ and $\omega_{0}\geq0.6U$.The lower limit for $\omega_{0}$ was determined by imposing the condition that $\sigma_{R}\geq0$,while all the other values were chosen in analogy with high temperature superconductors. The influence of doping (and its changes) on the conductivity can be studied through variations of the chemical potential which depends (among other parameters) on the band filling.A half-filled band ($n=1$) describes a chemically pure specimen. Positive deviations of the filling from the value $n=1$ experimentally correspond to doping with electron donors.Negative deviations describe the doping of a specimen by electron acceptors. 
\newpage 
The first test of the result obtained for the electrical conductivity was its application to the case of a 1D half-filled Hubbard model.Namely,it is known from rigorous theory that its conductivity is zero (for example \cite{HEINZ:95}).Inserting $n=1$ into eq.(29) and developing it in $t$ as a small parameter,it follows that the conductivity is approximately given by

\begin{equation}
\sigma_{R}(\omega)\cong10^{-7}(\frac{\omega_{P}}{\omega})^{2}t^{4} (4.56+.3\beta t)	
\end{equation}

Clearly,the numerical value of this expression is close to zero for physically acceptable values of the input parameters,of course excluding the case $\omega=0$.Looking from the experimental side,this can be interpreted as implying that weakly conducting phases of the Bechgaard salts can be described by a 1D Hubbard model with a small deviation of the band filling from the value of $1/2$. 

The experimental parameter which can be most easily controlled in experiments on organic conductors is the temperature of the specimen.The general conclusion of all such experiments is that the temperature dependence of the conductvity of the Bechgaard salts is extremely complex,and that it can not be explained by conventional theory of conductivity of metals.

In the calculations discussed in this chapter no attempt was made to reproduce the experimental conductivity of any particular salt.However,the idea was to determine whether or not the one band 1D Hubbard model can reproduce semiquantitatively the experimental data.Examples of real experimental data and theoretical calculations of the conductivity are avaliable in \cite{VE:00},\cite{CHE:97} and many other publications.

Equations (28) and (29) can be re-written as follows
\begin{equation}
\sigma_{R}(\omega_{0})=A\left[\omega_{0}^{2}-(bt)^{2}\right]^{-1}\left[Q+btZ(\omega_{0}+bt)^{-1}\right]	
\end{equation}
where the sumbols A,Q and Z denote various frequency independent functions which occur in the expression for the conductivity.Developing this result in $\omega_{0}$ up to second order terms,it follows that

\begin{equation}
\sigma_{R}(\omega_{0})\cong-A(Q+Z)(bt)^{-2}+AZ(bt)^{-3}(\omega_{0})-A(2Z-Q)(bt)^{-4}(\omega_{0})^{2}+...	
\end{equation}
Fitting this equation to measured frequency profiles of the conductivity,one could determine the functions A,Z,Q and $t$. An approximate value of the  static limit of the function $\sigma_{R}(\omega_{0})$  follows from eq.(32) as
\newpage
\begin{equation}
\sigma_{R}(\omega_{0}=0)\cong-A(Q+Z)(bt)^{-2}	
\end{equation} 

Fitting these two expressions to experimental data,it would be possible to determine the values of the functions A,Z,Q and the hopping integral $t$.Fixing all the parameters and varying the band filling would give the possibility of investigating the effects of doping on the conductivity.A preliminary investigation of this sort has been performed in \cite{CHE:97}.

\subsection{The equation of state and specific heat}

It is hoped that the reader has by now gained an idea about the complexity of theoretical studies of the Bechgaard salts.The aim of this sub-section is to outline a determination of the specific heat per particle under constant volume of a degenerate electron gas on a 1D lattice.Such a calculation may seem as an exercise in statistical physics and mathematics.It fact,in recent years measurements of specific heat and thermal transport properties of high $T_{c}$ superconductors and organic conductors have become a useful tool in studies of these systems.The literature in the field is steadily growing,but \cite{YU:92},\cite{JCL:99},\cite{KBEH:99} are some useful examples.

It will be assumed in the followin that the number of particles $N$ in the system is not conserved.Mathematically speaking,the starting point of the calculations is the equation

\begin{equation}
dG=-SdT+VdP+\mu dN	
\end{equation}
where $G=\mu N$ and all the symbols have 
their standard meanings.Inserting the definition of the thermodynamic potential $G$ into eq.(34) and differentiating with respect to $T$,one obtains 

\begin{equation}
V(1+\frac{1}{V}\frac{\partial V}{\partial T})\frac{\partial P}{\partial T}= S+\frac{\partial S}{\partial T}+\frac{\partial N}{\partial T}\frac{\partial \mu}{\partial T}+N\frac{\partial \mu}{\partial T}	
\end{equation}
In the last expression the obvious transformation $dP=\frac{\partial P}{\partial T}dT$ (and similar relations for other variables) was applied. 
Eq.(35) is expressed in terms of the bulk parameters of the system.In order to re-write it in terms of the local variables,a change of variables $N \rightarrow n V$ and $S \rightarrow n s V$ is necessary.After some algebra,it can be shown that 
\newpage
\begin{eqnarray}
V(1+\frac{1}{V}\frac{\partial V}{\partial T})\frac{\partial P}{\partial T}=	ns(V+\frac{\partial V}{\partial T})+V\frac{\partial n}{\partial T}(s+\frac{\partial\mu}{\partial T})+
\nonumber\\
nV\frac{\partial}{\partial T} (s+\mu)+n\frac{\partial \mu}{\partial T}\frac{\partial V}{\partial T}
\end{eqnarray}
 
We have here derived the differential equation of state (EOS) of any material.It is at first sight complicated,but in applications to the Bechgaard salts it can be considerably simplified.Experiments on these materials are almost always performed under fixed volume conditions \cite{ISH:90}.This implies that all terms in eq.(36) containing volume derivatives can be disregarded.The final form of the EOS of Q1D organic metals thus emerges as
\begin{equation}
\frac{\partial P}{\partial T}= (n+\frac{\partial n}{\partial T})(s+\frac{\partial \mu}{\partial T})+n\frac{\partial s}{\partial T}	
\end{equation}

The specific heat per particle is given by \cite{STAN:71}
\begin{equation}
c_{V}=\frac{T}{n}(\frac{\partial^{2}P}{\partial T^{2}})_{V}-T(\frac{\partial^{2}\mu}{\partial T^{2}})_{V}	
\end{equation}
 
Applying this definition and performing all the necessary algebra,one finaly gets the expression for the specific heat per particle 
\begin{eqnarray}
c_{V}=\frac{T}{n} (s+\frac{\partial\mu}{\partial T})(\frac{\partial n}{\partial T}+\frac{\partial^{2} n}{\partial T^{2}})+\frac{T}{n} \frac{\partial n}{\partial T}(2\frac{\partial s}{\partial T}+\frac{\partial^{2}\mu}{\partial T^{2}})+
\nonumber\\
T(\frac{\partial s}{\partial T}+\frac{\partial s^{2}}{\partial T^{2}})	
\end{eqnarray}
 
In order to apply eq.(39) to the degenerate electron gas on a 1D lattice,it is necessary to introduce into it appropriate expressions for the chemical potential,number density and entropy per particle.The chemical potential is given by eq.(16).As a first approximation,known results for the entropy per particle and the number density can be used \cite{CHE:98}.The entropy per particle is given by
\begin{equation}
s=\frac{Q}{n}\frac{\partial F_{3/2}(\mu/T)}{\partial T}	
\end{equation}
\newpage

The symbol $F_{3/2}(\mu/T)$ denotes a special case of a Fermi-Dirac integral \cite{CHE:98},$n$ is the electron number density and $Q$ is a combination of known constants (such as the electron mass and the Planck constant).Finally,the number density of a degenerate electron gas at low temperature can be expressed as \cite{CHE:98}

\begin{equation}
n \cong  A\times T^{15/2} \left[1+B\times T^{3/2}+...\right]	
\end{equation}
where the symbols $A$ and $B$ denote combinations of known constants.Inser-
ting eqs.(16),(40) and (41) into eq.(39) gives the final result for the specific heat of the electron gas on a 1D lattice. After all the necessary algebra,it would turn out to be highly non-linear,in line with the results of various experiments (such as \cite{JCL:99},\cite{BB:97}and later work).

\section{Conclusion}

The aim of this chapter was to review some results of theoretical studies of a class of organic conductors called the Bechgaard salts.
From the start it was prepared  as a "personal review",which simply means that it is to a large extent based on previous research results of the author.At the end of such a review,one can reflect for a moment on the possible future developement of research on the Bechgaard salts,and the organic conductors in general.Like Lord Kelvin in the $XIX$ century,one may be tempted to think that after more than 20 years since their discovery,most things about the Bechgaard salts are well known and only "small clouds" remain to be clarified.It seems (at least to the present author) that this would be totally wrong,and many interesting problems are waiting to be solved.A few of them are mentioned in the follow up.

The general question can be formulated as "Why should anybody on this planet study the organic conductors (and the Bechgaard  salts as a  paritular family of the organic conductors)?" The answer is "simple" : the Bechgaard salts are a Q1D example of strongly correlated systems,and therefore should be easier to understand than their 3D analogs,such as the high $T_{c}$ superconductors.However,in spite of all the efforts,the physical mechanism of high $T_{c}$ superconductivity has not yet been discovered.  

A basic problem is,of course,the applicability (or inapplicability) of the Fermi liquid model to these materials.Indications from various experiments (for example \cite{VE:00}) seem to be that the Fermi liquid is not a good description of the organic conductors.
To make things more interesting,there are also indications (such as \cite{CHE:97}or \cite{MO:98}) that reasonable agreement between calculations performed within the Fermi liquid model and experiments can be achieved in some cases.A definite solution of this dilemma is certainly one of the problems waiting to be solved as soon as possible. 

The calculations reported in the present chapter were performed within the "original version" of the memory function method.The method itself has been considerably reformulated and modernized in the mean time,
bringing it more "in line" with contemporary field theory.
For a recent application of a modernized version of the memory
function method see,for example,\cite{KUP:04}. Another method which may be useful for calculations like those described in this chapter(but also for the determination of the phase diagram) has been developed in \cite{SCWZ:93}.

The determination of the phase diagram of the Bechgaard salts is an interesting problem on its own.The temperature dependence of their conductivity can vary in some regions of the $P-T$ plane;accordingly,they can be insulators,superconductors but also normal conductors or semiconductors.The general form of the phase diagram of the Bechgaard salts is known for several decades.However,much more complicated is the possibility of determining the parameters (such as $t$) of a Bechgaard salt from the analysis of the phase boundaries on a $P-T$ diagram.For a recent example of such a work see \cite{CHE:04}. 

Interesting considerations have recently been made concerning the dimensionality of the Bechgaard salts.Namely,they are usually considered as one dimensional,or quasi one-dimensional.However,recent angular magnetoresistance osicllation (AMR) experiments \cite{KJO:03} on $(TMTSF)_{2}FSO_{3}$ have shown that a cylindrical Fermi surface can be formed for this material.Does this perhaps mean that the Bechgaard salts should be considered as Q2D materials is (to the knowledge of the present author) a completely open question.

This list of selected  problems concerning the Bechgaard salts could be continued,but it is hoped that the examples present illustrate sufficiently well that work on these materials is an interesting field of condensed matter physics.        

\section{Acknowledgement}
This review was prepared within the project 1231 financed by the Ministry of Science and Protection of the Environment of Serbia.

\end{document}